\documentstyle[preprint,revtex]{aps}

\begin{document}

\draft
\preprint{cond-mat/9312094}
\begin{title}
Erratum: Properties of Odd Gap Superconductors
\end{title}

\author{A. V. Balatsky}
\begin{instit}
 Theoretical Division\\ Los
Alamos National Laboratory, Los Alamos, NM 87545
\end{instit}
\author{Elihu Abrahams}
\begin{instit}
Serin Physics Laboratory, Rutgers University, P.O. Box
849, Piscataway, NJ 08855
\end{instit}
\author{D.J. Scalapino}
\begin{instit}
Department of Physics, University of California, Santa Barbara, CA
\end{instit}
\author {J.R. Schrieffer}
\begin{instit}
Department of Physics, Florida State University, Tallahassee, FL 32310
\end{instit}
\receipt{December 26 1993}
\begin{abstract}

Key words: superconductivity, odd frequency gap, composite operator.

\end{abstract}

\pacs{PACS Nos. 74.20-z;74.65+n}

In the recent paper \cite{Sandiego} an erroneous statement
that the composite order parameter for  odd-frequency pairing
was originally proposed by Berezinskii \cite{Ber} was
made.

Since the recent revival of the interest in
odd-frequency pairing \cite{BA},
the composite order parameter has been proposed in Abrahams et.al.
\cite{G4}, V. Emery and S. Kivelson \cite{EK} and in Coleman et.al.
\cite{C}.

\end{document}